\documentclass[aps,pra,twocolumn,showpacs]{revtex4-1}
\usepackage{graphicx}
\usepackage{amsmath}
\usepackage{amssymb}

\newcommand{\br}{{\bf r}}
\newcommand{\bk}{{\bf k}}
\newcommand{\bu}{{\bf u}}
\newcommand{\bv}{{\bf v}}

\begin{document}
\setlength{\arraycolsep}{1.5pt}
\bibliographystyle{prsty}

%Title of paper
\title{Elementary excitations in dipolar spin-1 Bose--Einstein condensates}
\author{J.~A.~M.~Huhtam\"aki} \email{jam@fyslab.hut.fi}
\author{P.~Kuopanportti}
\affiliation{Department of Applied Physics/COMP, Aalto
University, P.O. Box 14100, FI-00076 AALTO, Finland}

\begin{abstract}
We have numerically solved the low-energy excitation spectra of ferromagnetic Bose--Einstein condensates subject to dipolar interparticle interactions. The system is assumed to be harmonically confined by purely optical means, thereby maintaining the spin degree of freedom of the condensate order parameter. Using a zero-temperature spin-1 model, we solve the Bogoliubov excitations for different spin textures, including a spin-vortex state in the absence of external magnetic fields and a rapidly rotating polarized spin texture in a finite homogeneous field.
%The effect of dipolar interactions on the excitation energies is emphasized by considering a wide range of dipolar coupling strengths, beyond those of typical alkali-metal systems.
%In particular, we consider the effect of dipolar interactions on the excitation energies. Unlike density oscillations, the energies of magnetic quadrupole modes and spin waves characteristic of ferromagnetic spin-1 condensates are found to increase rapidly with the dipolar coupling strength.
In particular, we consider the effect of dipolar interactions on excitations characteristic of ferromagnetic condensates. The energies of spin waves and magnetic quadrupole modes are found to increase rapidly with the dipolar coupling strength, whereas the energies of density oscillations change only slightly.
%For the spin-vortex state, the dipolar interactions mix/lead to mixing of the density and spin waves.
\end{abstract}



\maketitle

\section{Introduction}
Energetically low-lying collective excitations are central for understanding systems that exhibit superfluidity. For example, sound waves are the lowest-energy excitations in superfluid ${}^{4} {\rm He}$, which explains why such a system may sustain frictionless flow below a critical velocity, namely, the speed of sound. Similarly, the excitation spectrum of a superconductor has a gap in the neighborhood of the Fermi energy, allowing electric currents to flow without resistance.

The study of collective modes in dilute atomic Bose--Einstein condensates (BECs) has had a key role in probing the physics of these versatile quantum systems. Condensates confined in strong magnetic traps are typically described by scalar-valued order-parameter fields, and their low-energy excitations are density oscillations. The possibility of trapping BECs optically has liberated the atomic spin degree of freedom, providing an opportunity for experimentalists to investigate condensates with spinor fields~\cite{Stamper-Kurn1998}. In the case of spin-1 alkali-metal BECs, such as ultracold vapors of ${}^{87}{\rm Rb}$ in its ground-state hyperfine manifold, the order parameter is a three-component vector~\cite{Ohmi1998,Ho1998}. The spin-1 gas of ${}^{87}{\rm Rb}$ is magnetized due to ferromagnetic interatomic spin coupling~\cite{Klausen2001}, and the related spin dynamics have been studied experimentally~\cite{Schmalljohann2004,Chang2004}. The lowest-lying collective excitations constitute a diverse set of density oscillations, spin waves, and quadrupolar spin fluctuations.

Recently, long-range anisotropic dipolar forces between the condensed bosons have attracted considerable interest. It has been theoretically predicted that dipolar interactions can give rise to interesting magnetic structures such as spin vortices~\cite{Yi2006,Kawaguchi2006,Huhtamaki2010} and helices~\cite{Kawaguchi2007,Huhtamaki2_2010}. Also, dipolar forces have been suggested to be involved in the disintegration of a helical spin texture into small magnetic domains in a condensate of spin-1 ${}^{87}{\rm Rb}$~\cite{Vengalattore2008}. In a similar system, dipolar interactions have been shown to induce magnetoroton softening in the excitation spectrum, linking the study to the physics of ${}^{4}{\rm He}$~\cite{Cherng2009}. In this article, we investigate how dipolar interactions affect the elementary excitation spectrum of a harmonically confined pancake-shaped spin-1 BEC. In particular, we solve the Bogoliubov spectra of a spin-vortex and a nearly spin-polarized (flare) state in the absence of polarizing external fields. In addition, we calculate the excitations of a dipolar BEC with the atomic spins polarized perpendicular to an external field. In a scalar dipolar system, collective excitations have been studied both theoretically~\cite{Ronen2006,Bijnen2010,Sapina2010} and experimentally~\cite{Bismut2010}.

The stationary states studied here using a spin-1 model have analogous counterparts in a semiclassical framework that describes condensates of particles with large dipole moments~\cite{Takahashi2006,Huhtamaki2010}. Thus, the presented results may shed light on the behavior of strongly dipolar systems such as the spin-3 condensate of ${}^{52}{\rm Cr}$~\cite{Griesmaier2005,Lahaye2009} which has atomic magnetic moments of $6\, \mu_\mathrm{B}$ and which has been produced by purely optical means~\cite{Beaufils2008}. Other strongly magnetic atomic gases that have been successfully cooled to ultralow temperatures include vapors of ${\rm Tm}$~\cite{Sukachev2010}, ${\rm Er}$~\cite{Berglund2008}, and ${\rm Dy}$~\cite{Lu2010}, which have atomic magnetic moments of $4\, \mu_\mathrm{B}$, $7\, \mu_\mathrm{B}$, and $10\, \mu_\mathrm{B}$, respectively.

\section{Theory}
In this work, we study spin-1 BECs using a zero-temperature %, such as ultracold gases of ${}^{87}{\rm Rb}$,
mean-field treatment, thereby neglecting possible effects due to the presence of a thermal component. The state of the condensate is described by a spinor wave function $\Psi(\br,t)=\left(\psi_1, \psi_0, \psi_{-1} \right)^T$ in the Zeeman basis. Its time dependence is governed by the Gross--Pitaevskii equation (GPE)
\begin{eqnarray}
\label{3DGP}
i\hbar \partial_t \Psi(\br,t) =\Big\{ \hat{h}^{3\mathrm{D}}(\br) &+& g_\mathrm{n}^{3\mathrm{D}} n(\br) + g_\mathrm{s}^{3\mathrm{D}} \sum_\alpha M_\alpha(\br)\hat{S}_\alpha \nonumber \\
&+& g_\mathrm{d}^{3\mathrm{D}} \sum_\alpha I_\alpha(\br) \hat{S}_\alpha \Big\} \Psi(\br,t).
\end{eqnarray}
Here, $\hat{h}^{3\mathrm{D}}=-\hbar^2 \nabla^2/2m+V_{\rm trap}(\br)-p\hat{S}_z+q\hat{S}_z^2$ is the single-particle Hamiltonian with the optical trapping potential $V_{\rm trap}=m \left[ \omega_r^2 (x^2+y^2)+\omega_z^2 z^2 \right]/2$. The last two terms of $\hat{h}^{3\mathrm{D}}$ account for the linear and quadratic Zeeman terms with the external homogeneous magnetic field pointing along the $z$ axis. The constant $p$ depends linearly and $q$ quadratically on the strength of the applied field. In Eq.~(\ref{3DGP}), $n(\br)=\Psi^\dagger \Psi = \sum_{k=-1}^1 \psi_k^* \psi_k$ is the particle density and $M_\alpha(\br)=\Psi^\dagger \hat{S}_\alpha \Psi$, $\alpha\in\left\{x,y,z\right\}$, stands for the $\alpha$th component of the local magnetization with $\{\hat{S}_\alpha\}$ denoting the standard spin-1 operators. Furthermore, $I_\alpha(\br)$ is a spin-dependent effective potential arising from nonlocal dipolar interactions (see Appendix). The coupling constants determining the strengths of the density--density, spin--spin, and dipole--dipole interactions are given by $g_\mathrm{n}^{3\mathrm{D}}=4\pi \hbar^2 (a_0+2a_2)/3m$, $g_\mathrm{s}^{3\mathrm{D}}=4\pi \hbar^2 (a_2-a_0)/3m$, and $g_\mathrm{d}^{3\mathrm{D}}=\mu_0 \mu_\mathrm{B}^2 g_\mathrm{F}^2/4\pi$, respectively. Here, $\mu_0$ denotes the permittivity of the vacuum, $\mu_\mathrm{B}$ is the Bohr magneton, and $g_\mathrm{F}$ is the Land\'e factor ($g_\mathrm{F}=-1/2$ for ${}^{87}\mathrm{Rb}$ with $F=1$).

From now on, we will assume that the components of the order parameter have the same Gaussian profile along the $z$ axis, $\Psi(\br)=\Psi(x,y) A_\sigma \exp(-z^2/2\sigma^2)$ with $A_\sigma=\sigma^{-1/2} \pi^{-1/4}$, implying that the direction of magnetization does not depend on $z$. This approximation should be accurate in the limit of tight axial trapping, $\omega_z \gg \omega_r$. In the noninteracting limit, the ansatz produces the ground state if $\sigma=a_z \equiv \sqrt{\hbar/m\omega_z}$. Thus, in the case of a pancake-shaped BEC, we may reduce the GPE to the effectively two-dimensional form
\begin{eqnarray}
\label{2DGP}
i\hbar \partial_t \Psi(x,y,t) =\Big\{ \hat{h} &+& g_\mathrm{n} n + g_\mathrm{s} \sum_\alpha M_\alpha \hat{S}_\alpha \nonumber \\
&+& g_\mathrm{d} \sum_\alpha I^\sigma_\alpha \hat{S}_\alpha \Big\} \Psi(x,y,t),
\end{eqnarray}
where the effective potentials are now evaluated at $z=0$, $\hat{h}=-\hbar^2 \left(\partial_x^2 + \partial_y^2 \right)/2m+m\omega_r^2 \left(x^2+y^2\right)/2-p\hat{S}_z+q\hat{S}_z^2$, and $g_\mathrm{i}=g_\mathrm{i}^{3\mathrm{D}}/\sqrt{2\pi}\sigma$, $\mathrm{i}\in \{\mathrm{n},\mathrm{s},\mathrm{d} \}$. The dipolar integrals $\{I^\sigma_\alpha(x,y)\}$ include the width $\sigma$ of the Gaussian profile as a parameter. The precise forms of these functions are derived in the Appendix.

Stationary states are obtained with the ansatz $\Psi(x,y,t)=\Psi(x,y) \exp \left( -i\mu t/\hbar \right)$, where $\mu$ is the chemical potential determined by normalizing the wave function to the total particle number $N=\int d{\bf{r}} \Psi^\dagger \Psi$. In the presence of strong axial confinement, excitations of the condensate are frozen in the axial direction. Small-amplitude oscillations about a stationary state $\Psi(x,y)$ are found by assuming the wave function has the form 
\begin{equation}
\Psi(x,y,t)=e^{-i\mu t/\hbar} \left[ \Psi(x,y)+\bu(x,y) e^{-i\omega t} + \bv^\ast(x,y) e^{i\omega^* t}\right],
\end{equation}
where the vectors $\bu$ and $\bv$ are called quasiparticle amplitudes. Substituting this trial into Eq.~(\ref{2DGP}) leads to the Bogoliubov equations of the general form
\begin{equation}
\label{BdG}
\begin{pmatrix} \hat{\mathcal{A}} & \hat{\mathcal{B}} \\
-\hat{\mathcal{B}}^* & -\hat{\mathcal{A}}^* \end{pmatrix} \begin{pmatrix} \bu \\ \bv \end{pmatrix} = \hbar \omega \begin{pmatrix} \bu \\ \bv \end{pmatrix}.
\end{equation}
Here, $\hbar \omega$ denotes the excitation energy of the mode $\{ \bu, \bv \}$, and $\hat{\mathcal{A}}$ and $\hat{\mathcal{B}}$ are linear operators whose detailed forms are given in the Appendix.

An external magnetic field ${\bf B}$ exerts a torque ${\bf M} \times {\bf B}$ on a magnetic moment ${\bf M}$, which causes the moment to precess about the direction of the magnetic field with the Larmor frequency $\omega_\mathrm{L}$. Thus, in the case of a homogeneous axial field $B_z {\bf e}_z$, the magnetization in a stationary state is polarized along the $z$ axis. Yet, at least in the absence of dipolar interactions, we can allow for more general magnetization textures by considering states that are stationary in a spin reference frame that rotates about the magnetic field. Such states are of the form  $\Psi_p=\exp(ip\hat{S}_z t/\hbar) \Psi_0$, where $p/\hbar$ is the precession frequency about $\hat{S}_z$ and $\Psi_0$ is a stationary solution of Eq.~(\ref{2DGP}) with $p=0$. This simple picture holds when the dipolar interactions are absent, but when $g_\mathrm{d}\neq 0$, the dipolar potential is time dependent in the rotating frame and the analysis becomes more involved. However, for systems considered in this work, the Larmor frequency in a typical magnetic field of strength \mbox{$100$ $\mathrm{m} \mathrm{G}$} is in the $\mathrm{M} \mathrm{Hz}$ regime, which is much faster than other relevant dynamics of the condensate, such as spin and density oscillations whose characteristic frequencies are in the $\mathrm{k} \mathrm{Hz}$ regime. Therefore, the system can be viewed as being subject to an effective dipolar potential obtained by time-averaging the instantaneous dipole--dipole potential over one Larmor cycle~\cite{Kawaguchi2007}. The time-averaged potentials are given in the Appendix.

%An external magnetic field ${\bf B}$ exerts a torque ${\bf \Gamma}={\bf M} \times {\bf B}$ on a magnetic moment ${\bf M}$, which causes the magnetic moment to precess about the direction of the magnetic field with the Larmor frequency $\omega_\mathrm{L}$. Thus, in the case of a homogeneous axial magnetic field $B_z {\bf e}_z$, the magnetization in a stationary state is polarized along the $z$ axis. However, in the absence of dipolar interactions, $g_\mathrm{d}=0$, a stationary solution $\Psi_0$ to Eq.~(\ref{2DGP}) with $p=0$ may be viewed as a stationary state in a frame rotating about $\hat{S}_z$ with the frequency $p$, which is evident from the fact that $\Psi_p=\exp(ip\hat{S}_z t/\hbar) \Psi_0$ satisfies Eq.~(\ref{2DGP}) [THIS SENTENCE IS VERY HARD TO GRASP]. With finite dipolar interactions, $g_\mathrm{d}>0$, the dipolar potential becomes time dependent in the rotating frame of reference. However, for systems considered in this work, the Larmor frequency in a typical magnetic field of strength \mbox{$100$ $\mathrm{m} \mathrm{G}$} is in the $\mathrm{M} \mathrm{Hz}$ regime, which is much faster than other relevant dynamics of the condensate, such as spin and density oscillations  whose characteristic frequencies are in the $\mathrm{k} \mathrm{Hz}$ regime. Therefore, the system can be viewed as being subject to an effective dipolar potential obtained by time-averaging the instantaneous dipole--dipole potential over one Larmor cycle. The time-averaged potentials are given in the Appendix.

In the numerics, we discretize the effectively two-dimensional equations on a uniform grid. The Bogoliubov matrix in Eq.~(\ref{BdG}) is typically full when $g_\mathrm{d}>0$ due to nonlocality of the dipolar potentials, and explicit diagonalization of the large eigensystem becomes cumbersome. Instead, we use the implicitly restarted Arnoldi iteration implemented in the ARPACK library for solving only the relevant part of the spectrum, namely, the energetically low-lying eigenmodes.
%The technique relies on the repetitive operation by the system matrix with the benefit that the multiplication by the matrix can be viewed simply as a linear operation without the need for storing the large matrix in the memory.
The method relies on repetitive operation by the system matrix without the need to store it in memory.
The two-dimensional Fourier transforms used in evaluating the dipolar potentials are implemented efficiently with a fast Fourier transform (FFT).

In the following, we choose $a_r=\sqrt{\hbar/m\omega_r}$ and $\hbar \omega_r$ for the units of length and energy, respectively. We set the dimensionless coupling constant to $\tilde{g}^{3\mathrm{D}}_\mathrm{n}=g^{3\mathrm{D}}_\mathrm{n} N/\hbar\omega_r a_r^3=500$, which corresponds roughly to $10^4$ ${}^{87}\mathrm{Rb}$ atoms in a trap with $\omega_r/2\pi=100$ $\mathrm{Hz}$. Moreover, we choose $\sigma=\sqrt{2}/5$, yielding $\tilde{g}_{\mathrm{n}}=\tilde{g}^{3\mathrm{D}}_\mathrm{n}/\sqrt{2\pi}\sigma  \approx 700$. The condensate is assumed to be ferromagnetic with $g_{\mathrm{s}}/g_{\mathrm{n}}=-0.01$. Qualitatively, the results we present are not very sensitive to the values of $g_\mathrm{n}$, $g_\mathrm{s}$, or $\sigma$. Instead of fixing the dipolar coupling strength to some
predetermined value (for ${}^{87}\mathrm{Rb}$, $g_{\mathrm{d}}/g_{\mathrm{n}}=4.2 \times 10^{-4}$), we consider a range of interaction strengths within the entire stability window $0 \leq g_{\mathrm{d}}/g_{\mathrm{n}} \leq 0.24$~\cite{Yi2004,Huhtamaki2010}.
%Hence, the study may also provide useful information for strongly dipolar systems, such as gases of ${}^{52} \mathrm{Cr}$~\cite{Lahaye2009}.
Although alkali-metal condensates are typically subject to weak dipolar forces, it might be feasible to tune the ratio $g_{\mathrm{d}}/g_{\mathrm{n}}$ experimentally by utilizing optical Feshbach resonances~\cite{Fedichev1996,Theis2004}. For the external magnetic field, we choose ${\bf B}=B_z {\bf e}_z$ with $B_z = 100$ $\mathrm{mG}$. In such a field, the Larmor frequency of a ${}^{87}{\rm Rb}$ atom is $\omega_\mathrm{L} = g_\mathrm{F}\mu_\mathrm{B} B_z/\hbar \approx 2\pi \times 70$~${\rm kHz}$, and the quadratic Zeeman shift has a magnitude of $q/\hbar \approx 2\pi \times 0.70$ ${\rm Hz}$~\cite{Vengalattore2010}.

\begin{figure}[!t]
\includegraphics[width=230pt]{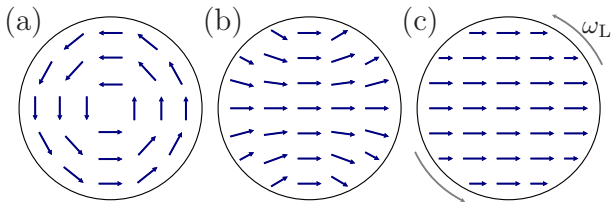}
\caption{\label{FIG1} (Color online) Schematics of the three spin textures considered in this work: (a) \emph{spin vortex}, (b) \emph{flare}, and (c) \emph{spin-polarized state}. The spin vortex and flare are studied in the absence of magnetic fields, whereas the spin in the polarized state is precessing rapidly with the Larmor frequency $\omega_\mathrm{L}$ about a homogeneous field perpendicular to the plane of the figure.}
\end{figure}

\begin{figure}[!t]
\includegraphics[width=230pt]{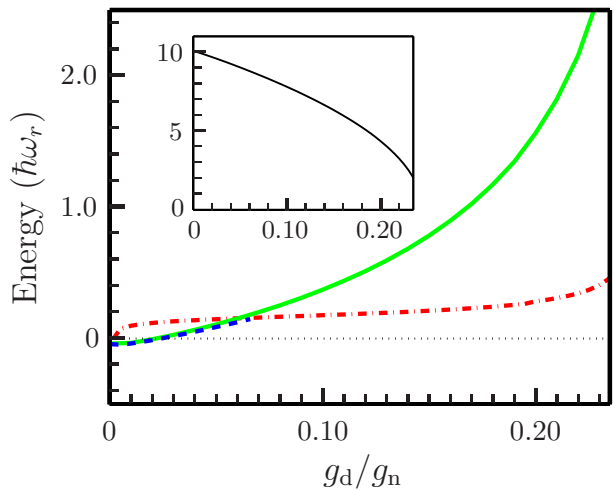}
\caption{\label{FIG2} (Color online) Total energies of the spin-polarized state in an external axial magnetic field (solid line) and the flare state (dashed line) as a function of the dipolar coupling strength $g_\mathrm{d}$. The energies are given with respect to the total energy of the spin-vortex state shown in the inset and by the dotted line in the main figure. The Mermin--Ho vortex (dash-dotted line) is an excited state for all parameter values considered.}
\end{figure}

\begin{figure*}[!t]
\includegraphics[width=500pt]{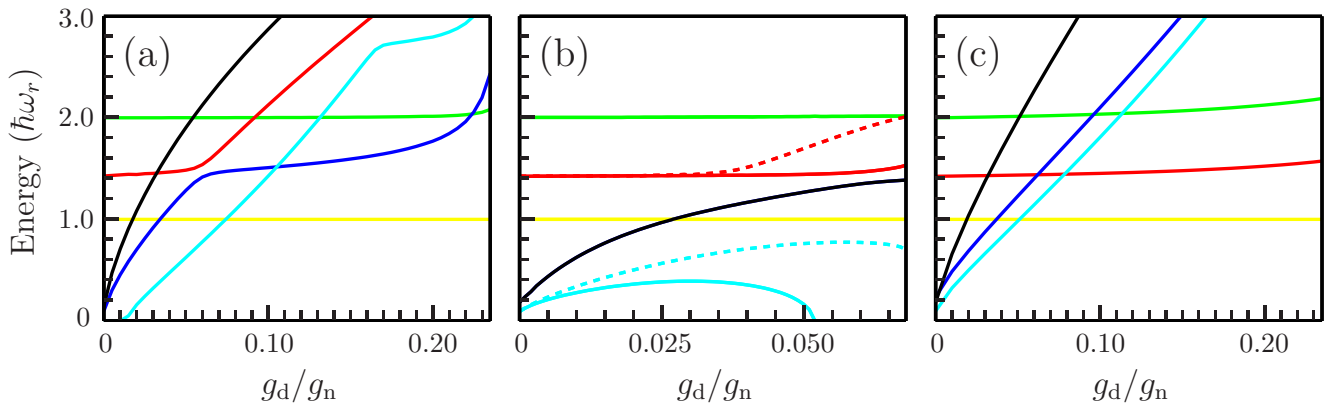}
\caption{\label{FIG3} (Color online) Excitation energies of selected Bogoliubov modes as a function of the dipolar coupling strength $g_\mathrm{d}$. The graphs refer to the (a) spin vortex, (b) flare, and (c) spin-polarized state illustrated in Fig.~\ref{FIG1}. %, and whose energies are shown in Fig.~\ref{FIG2}.
In the limit $g_\mathrm{d}/g_\mathrm{n} \to 0$, the three highest-energy modes in each figure are the density breathing ($\kappa=0$), quadrupole ($\kappa=\pm 2$), and dipole ($\kappa=\pm 1$) modes, in order of descending energy. The three lower-lying modes are the lowest magnetic quadrupole oscillation ($\kappa=0$) and spin waves ($\kappa=\pm 2$ and $\kappa=\pm 1$). The energies of the magnetic quadrupole mode and the spin waves increase rapidly with dipolar coupling strength, whereas the energies of the density oscillations are less affected. The density and spin oscillations become  mixed due to the dipolar forces. In the spin-vortex case, the density and spin oscillations of similar symmetry exhibit avoided crossings.}
%The dipolar interaction breaks the cylindrical symmery in the flare state lifting the degeneracy of the excitations (dashed curves).
\end{figure*}

\section{Results}
The homogeneous spin-polarized spin-1 system has a rich variety of excitations~\cite{Ho1998,Ohmi1998}: The collective density oscillations obey the typical Bogoliubov dispersion relation $\varepsilon(k) = \sqrt{\varepsilon^2_0(k)+2g_\mathrm{n} n \varepsilon_0(k)}$, where $\varepsilon_0(k)=\hbar^2 k^2 /2m$. The long-wavelength density modes have a linear phonon-like spectrum and the short-wavelength excitations approach the quadratic free-particle case. The spin waves have a free-particle-like spectrum $\varepsilon(k)=\varepsilon_0(k)$. The magnetic quadrupole oscillations are periodic modulations in the quadrupole tensor $Q_{\alpha \beta} = 2\delta_{\alpha \beta}/3-\langle \hat{S}_\alpha \hat{S}_\beta + \hat{S}_\beta \hat{S}_\alpha \rangle/(2 n)$ that describe local spin fluctuations~\cite{Mueller2004}. They have a quadratic dispersion with an energy gap, $\varepsilon(k)=\varepsilon_0(k)+2|g_\mathrm{s}|n$. Here, we consider these excitations in an inhomogeneous system in the presence of dipolar interparticle forces.

We have solved the low-lying excitations in a ferromagnetic dipolar spin-1 BEC for the three types of stationary states illustrated in Fig.~\ref{FIG1}. First, we consider two stationary states in the absence of an external polarizing field;namely, the spin vortex~[Fig.~\ref{FIG1}(a)] and the flare~[Fig.~\ref{FIG1}(b)], studied previously in Refs.~\cite{Takahashi2006,Yi2006,Kawaguchi2006,Huhtamaki2010}. We choose the spin-vortex order-parameter field to be of the form $\psi_k(x,y)=\psi_k(r) \exp(-i k \varphi)$. In such a state, the transverse magnetization rotates through an angle of $2\pi$ about the unmagnetized vortex core. Second, we consider the ground state with vanishing axial magnetization in a homogeneous axial magnetic field. Due to the rapid Larmor precession of the transverse magnetization, the condensed particles feel a time-averaged dipolar potential, which results in a transversally spin-polarized texture~[Fig.~\ref{FIG1}(c)]. The total energies of these stationary states as a function of the dipolar coupling strength $g_\mathrm{d}$ are depicted in Fig.~\ref{FIG2}. The inset shows the energy of the spin-vortex state, which is the ground state for $g_\mathrm{d}/g_\mathrm{n} \ge 0.024$. %The solid and dashed lines in the main graph represent the energies of the spin-polarized and flare states with respect to the energy of the spin vortex state.
For comparison, we also show the energy of the Mermin--Ho vortex that carries both mass and spin currents.

The spin vortex and the spin-polarized state are cylindrically symmetric, and therefore their excitations can be written in the form
\begin{eqnarray}
{\bf u}_\kappa(r,\varphi)&=&e^{i \left(\kappa-s \hat{S}_z \right)\varphi} {\bf u}_\mathrm{\kappa}(r),\\
{\bf v}_\kappa(r,\varphi)&=&e^{i \left(\kappa+s \hat{S}_z \right)\varphi} {\bf v}_\mathrm{\kappa}(r).
\end{eqnarray}
We have chosen the spin winding $s=1$ for the spin vortex and $s=0$ for the spin-polarized state. The quantum number $\kappa \in \mathbb{Z}$ determines the phase winding of the excitation. We are interested in low-energy excitations which typically have small values of $\kappa$, such as breathing ($\kappa=0$), dipole ($\kappa=\pm 1$), and quadrupole ($\kappa=\pm 2$) modes. For fixed $\kappa$, there are several low-energy excitations depending on the nature of the mode (density wave, spin wave, or magnetic quadrupole wave) and the number of radial nodes in the components of ${\bf u}_\mathrm{\kappa}(r)$ and ${\bf v}_\mathrm{\kappa}(r)$.

%\subsection{Excitation Energies}
Figure~\ref{FIG3} shows the dependence of the excitation energies on the dipolar coupling strength $g_\mathrm{d}$. In the limit of vanishing $g_\mathrm{d}$, the low-lying parts of the spectra host myriads of excitations, of the order of a hundred in the range $\varepsilon \in [0, 3\,\hbar \omega_r]$ for each state. However, in the limit of strong dipolar interactions, only a few excitations remain within this energy window. For clarity, we have presented the energies only for the most interesting modes.

In Fig.~\ref{FIG3}(a), the three highest-lying modes in the limit $g_\mathrm{d}/g_\mathrm{n} \to 0$ are the density breathing, quadrupole, and dipole modes with energies $\left( 2,\sqrt{2},1 \right)$ $\hbar \omega_r$ and quantum numbers $\kappa=0, \pm 2, \pm 1$, respectively. The excitation energy of the dipole mode is determined solely by the external trapping potential, and hence it is independent of internal interactions, such as dipolar forces. In general, excitations that induce strong density fluctuations are relatively robust against changes in the dipolar coupling strength. The three remaining modes shown, in order of decreasing energy, are the lowest magnetic quadrupole oscillation ($\kappa=0$), the quadrupole spin wave ($\kappa=\pm 2$), and the core-localized mode ($\kappa=\pm 1$). The energies of these excitations increase rapidly with increasing $g_\mathrm{d}/g_\mathrm{n}$. %, as if such collective degrees of freedom were frozen due to the dipolar forces.
The core-localized excitation corresponds to an imaginary eigenvalue for $g_\mathrm{d}/g_\mathrm{n} \lesssim 0.01$, which signifies the dynamical instability of the spin vortex.

The spin-vortex state $\psi_k(x,y)=\psi_k(r) \exp(-ik\varphi)$ is invariant under the discrete symmetry $\psi_k \to \psi^*_{-k}$. Excitations without this invariance are doubly degenerate. In Fig.~\ref{FIG3}(a), these include all but the density breathing mode and the lowest magnetic quadrupole oscillation. For the Mermin--Ho vortex $\psi_k(x,y)=\psi_k(r) \exp[-i(k+1)\varphi]$, this symmetry is broken, and hence the spectrum is mostly nondegenerate. The two core-localized modes of the Mermin--Ho vortex have negative excitation energies for weak dipolar interactions, $g_\mathrm{d}/g_\mathrm{n} \lesssim 0.1$, which implies that the Mermin--Ho vortex is locally energetically unstable.

%The polar-core spin-vortex state $\psi_k(x,y)=\psi_k(r) e^{-ik\varphi}$ is invariant under the continuous symmetry transformation of a combined spatial and spin rotation. Moreover, the state has the discrete symmetry $\psi_k \to \psi^*_{-k}$. These properties are inherited by the excitations. The continous symmetry implies that the quasiparticle amplitudes have the forms ${\bf u}_{\mathrm{n},\kappa}(r,\varphi)=e^{-i \kappa \varphi \hat{S}_z} {\bf u}_\mathrm{n}(r)$ (similarly for ${\bf v}$ with $\kappa \to \kappa+2$), where $\mathrm{n} \in \mathbb{N}$ and $\kappa \in \mathbb{Z}$ indicate the number of radial nodes and the spin winding of the excitation, respectively.

The flare state is only found within the interval $0 \le g_\mathrm{d}/g_\mathrm{n} \lesssim 0.07$, beyond which we obtain a state with the same symmetry but hosting two spin vortices. Energies of several excitations of the flare state are shown in Fig.~\ref{FIG3}(b). The three highest-lying excitations in the limit $g_\mathrm{d}/g_\mathrm{n} \to 0$ are the density breathing, quadrupole, and dipole modes. The two modes below them are the lowest magnetic quadrupole oscillation and the dipole spin wave. The dipolar interaction breaks the rotational symmetry of the flare state, which lifts the degeneracy of all but the density dipole mode. Previously, the frequency splitting of the counter-rotating density quadrupole modes due to the presence of a mass vortex in a scalar system was utilized to measure the angular momentum of a BEC~\cite{Zambelli1998,Chevy2000}. In the absence of external magnetic fields, the lifted degeneracy in Fig.~\ref{FIG3}(b) could provide a spectroscopic means of measuring the strength of the dipolar forces. For $g_\mathrm{d}/g_\mathrm{n} \gtrsim 0.05$, the lower branch of the dipole spin wave turns imaginary.

In Fig.~\ref{FIG3}(c), the excitation energies for the transversally spin-polarized state in a homogeneous axial field are shown analogously to the previous cases. The three lowest-lying modes in the limit $g_\mathrm{d}/g_\mathrm{n} \to 0$ are the lowest magnetic quadrupole oscillation ($\kappa=0$) and two lowest spin waves ($\kappa=\pm 2$ and $\kappa=\pm 1$).%The most notable difference in this case compared to the spin vortex state is the decoupling of the spin- and density waves, and in particular the absence of the avoided crossings.

The avoided crossings of the excitation energies
%with respect to the dipolar coupling strength
occur for density and spin waves of the same quantum number $\kappa$; for example,
such a crossing is visible in Fig.~\ref{FIG3}(a) for the quadrupolar
density and spin oscillations with $\kappa=\pm 2$ at $g_\mathrm{d} / g_\mathrm{n} \approx 0.06$. A similar phenomenon has been previously studied in anisotropically trapped scalar BECs~\cite{Reidl1999,Woo2005}. In our case, the perturbing Hamiltonian is the dipolar interaction term. The interaction couples the neighboring quantum numbers $\kappa$ in the spin-vortex case but does not have an azimuthal dependence in the rapidly precessing spin-polarized state. Therefore, the avoided crossings appear for the spin vortex in Fig.~\ref{FIG3}(a), as well as for the Mermin--Ho vortex, but are absent in the spin-polarized case.

\begin{figure}[!t]
\includegraphics[width=230pt]{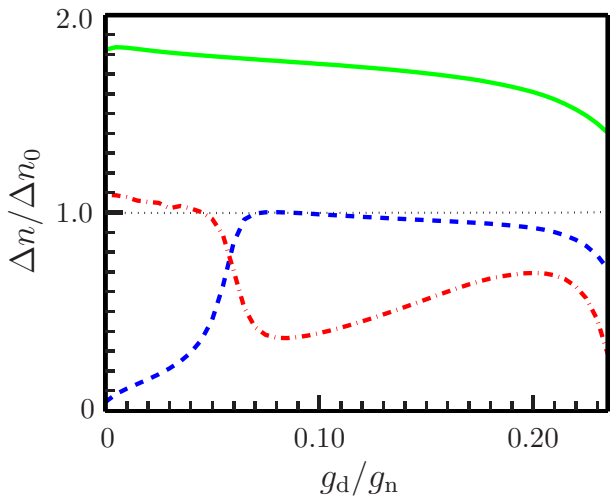}
\caption{\label{FIG4} (Color online) Condensate density modulation $\Delta n$ caused by a given excitation of the spin vortex compared to that of a simple translation $\Delta n_0$. In the limit $g_\mathrm{d}/g_\mathrm{n} \to 0$, the curves show the modulations due to the density breathing (solid), density quadrupole (dashed), density dipole (dotted), and the quadrupole spin (dash-dotted) modes. The roles of the quadrupole density and spin waves are interchanged in the neighborhood of the avoided crossing shown in Fig.~\ref{FIG3}(a).}
\end{figure}

In the absence of dipolar interactions, it is usually straightforward to distinguish the density waves from spin waves in the excitation spectra. The long-range anisotropic dipolar forces naturally couple spin and density oscillations. In order to quantify the change in the density due to a given excitation, we define the time-averaged integrated density modulation~\footnote{Analogously, we could define a modulation in the magnetization of the condensate related to a given excitation. This results in a similar but inverted graph.}
\begin{equation}
\Delta n = \int d\br \big| \langle \Psi'^\dagger \Psi'-\Psi^\dagger \Psi \rangle \big|= \frac{4}{\pi} \epsilon \int d\br |\Psi^\dagger \bu+\Psi^T \bv|.
\end{equation}
Here, $\langle \cdot \rangle$ denotes the time average, $\Psi$ is the stationary state, and $\Psi'=\Psi+\epsilon \left( \bu e^{-i\omega t}+\bv^* e^{i\omega^* t} \right)$ is the excited state ($\epsilon \ll 1$). We normalize $\Delta n$ with respect to the density modulation caused by a simple translation of the condensate, $\Delta n_0 = \int d\br \big| \tilde \Psi^\dagger \tilde \Psi-\Psi^\dagger \Psi \big|$, where $\tilde \Psi(\br) = \Psi(\br+\epsilon {\bf e})$ and ${\bf e}$ is an arbitrary unit vector in the $xy$ plane. According to this definition, $\Delta n/ \Delta n_0 = 1$ for the dipole mode, since it represents merely a translation of the condensate. In Fig.~\ref{FIG4}, this is shown as a dotted line for reference. The solid, dashed, and dash-dotted lines represent the ratio $\Delta n/ \Delta n_0$ for the density breathing, density quadrupole, and spin quadrupole modes, respectively. In the limit $g_\mathrm{d} / g_\mathrm{n} \to 0$, only the density oscillations impart significant density modulations. However, as the dipolar coupling increases, even the spin waves start yielding density modulations. Intersection of the quadrupolar density- and spin-wave curves clearly demonstrates that their roles are interchanged in the vicinity of the avoided crossing discussed in relation to Fig.~\ref{FIG3}(a).

%Due to partial cancellation of dipolar forces as a result of the rapid Larmor precession, the state is found even beyond the limit of dipolar collapse, $g_\mathrm{d}/g_\mathrm{n} \approx 1/4$.

\section{Discussion}
In summary, we have numerically studied the collective excitations of a harmonically confined dipolar spin-1 condensate. The Bogoliubov quasiparticle spectra were solved for three types of spin configurations often encountered in ferromagnetic systems: the spin vortex, the flare, and the spin-polarized texture precessing rapidly due to a homogeneous external magnetic field. Particular emphasis was placed on investigating the effect of dipole-dipole interactions on the quasiparticle spectra: the excitation energies of spin waves and magnetic quadrupole oscillations were found to increase rapidly as a function of the dipolar coupling strength, whereas the energies of excitations related to collective density oscillations were essentially unaffected.

As an extension of the present work, one could solve the excitations of dipolar pancake-shaped condensates in a wider energy range in order to study the effect of trapping on the roton-maxon spectrum. Previously, the emergence of the roton minimum has been investigated in the quasi-two-dimensional limit both in scalar~\cite{Santos2003} and spin-1 systems~\cite{Cherng2009}. Also, taking the dipolar forces into account in studying the dynamical instability of the spin helix of Ref.~\cite{Cherng2008} might be relevant due to their possible stabilizing effect~\cite{Huhtamaki2_2010}.

It would be interesting to investigate the collective excitations at nonzero temperatures. However, based on earlier studies of scalar condensates~\cite{Hutchinson1997,Ronen2007}, we expect that finite-temperature effects do not alter the qualitative features of the excitation spectra at temperatures at which experiments with dilute BECs are typically carried out.

\begin{acknowledgments}
The authors thank the Academy of Finland, the Emil Aaltonen Foundation, the KAUTE Foundation, and the V\"ais\"al\"a Foundation for financial support. M.~M\"ott\"onen, V.~Pietil\"a, and T.~P.~Simula are acknowledged for useful discussions.
\end{acknowledgments}

\appendix*
\section{Effective dipolar potentials}
Here, we briefly present the steps required for deriving the effectively two-dimensional GPE, Eq.~(\ref{2DGP}), and calculate the dipolar integrals $\{I_\alpha^\sigma(x,y) \}$ therein. Repeated indices are to be summed over throughout the Appendix. All but the last term in Eq.~(\ref{2DGP}) are obtained in a straightforward manner by substituting the ansatz $\Psi(\br)=\Psi(x,y) G(z)$ with $G(z)=A_\sigma \exp(-z^2/2\sigma^2)$ and $A_\sigma=\sigma^{-1/2} \pi^{-1/4}$ into Eq.~(\ref{3DGP}), multiplying by $G(z)$, and integrating over $z$.  The dipolar integrals in Eq.~(\ref{3DGP}) read
\begin{equation}
\label{3DI}
I_\alpha(\br)=\int d\br' D_{\alpha \beta}(\br-\br') M_\beta(\br'),
\end{equation}
with $D_{\alpha \beta}(\br)=\left(\delta_{\alpha \beta}r^2 - 3 r_\alpha r_\beta \right)/r^5$. After the aforementioned substitution procedure, the dipolar term becomes $g_\mathrm{d} I^\sigma_\alpha(x,y) \hat{S}_\alpha \Psi(x,y)$, where
\begin{equation}
I^\sigma_\alpha(x,y)=\frac{1}{A_\sigma} \int d\br' D_{\alpha \beta}(\br-\br') M_\beta(x',y') G(z') \Big|_{z=0}.
\end{equation}
The right-hand side is easily simplified with the convolution theorem. The required Fourier transforms are given by~\footnote{We use the nonunitary convention with a factor of $1/(2\pi)^d$ in the inverse Fourier transformation.}
\begin{eqnarray}
\hat{\mathcal{F}} \left\{ D_{\alpha \beta}(\br)\right\} &=& -\frac{4\pi}{3} \left(\delta_{\alpha \beta} -3\frac{k_\alpha k_\beta}{k^2} \right)=K_{\alpha \beta}(\bk),\\
\hat{\mathcal{F}} \left\{ G(z) \right\} &=& \frac{\sqrt{2}}{A_\sigma} e^{-(k_z \sigma)^2/2}.
\end{eqnarray}
By denoting $\tilde{M}_\beta = \hat{\mathcal{F}} \{ M_\beta \}$, we obtain
\begin{eqnarray}
I^\sigma_\alpha(x,y)=\sqrt{2\pi}\sigma \hat{\mathcal{F}}^{-1} \Big\{K_{\alpha \beta}(k_x,k_y,k_z) \nonumber \\
\times \tilde{M}_{\beta}(k_x,k_y) e^{-(k_z \sigma)^2/2} \Big\}\bigg|_{z=0}.
\end{eqnarray}
Because the expression is evaluated at $z=0$, the inverse Fourier transform over $k_z$ simply reduces to an integral. By evaluating the integrals and denoting
\begin{eqnarray}
K'_{xy} &=& \frac{k_x k_y}{k_r^2} f(k_r \sigma)=K'_{yx},\\
K'_{xx} &=& -\frac{4\pi}{3}+\frac{k_x^2}{k_r^2} f(k_r \sigma),\\
K'_{yy} &=& -\frac{4\pi}{3}+\frac{k_y^2}{k_r^2} f(k_r \sigma),\\
K'_{zz} &=& \frac{8\pi}{3}-f(k_r \sigma), \\
K'_{\alpha \beta}&=&0 \quad \quad \rm{otherwise},
\end{eqnarray}
where $f(\xi)=\left(2\pi\right)^{3/2}\xi \exp\left( \xi^2/2 \right) \mathrm{erfc}\left( \xi/\sqrt{2} \right)$ and $k_r=\sqrt{k_x^2+k_y^2}$, we find
\begin{equation}
\label{appendixI}
I^\sigma_\alpha(x,y)=\hat{\mathcal{F}}^{-1} \Big\{ K'_{\alpha \beta}(k_x,k_y) \tilde{M}_{\beta}(k_x,k_y) \Big\}.
\end{equation}
Hence, within the assumption of the Gaussian axial profile, updating the dipolar potentials for a given state $\Psi(x,y)$ requires the evaluation of three two-dimensional Fourier and inverse Fourier transforms. By defining the linear operator $\hat{\mathcal{L}}_{\alpha \beta} \{ \cdot \} = \hat{\mathcal{F}}^{-1} \big\{ K'_{\alpha \beta} \hat{\mathcal{F}} \{ \cdot \} \big\}$, Eq.~(\ref{appendixI}) can be written simply as $I^\sigma_\alpha=\hat{\mathcal{L}}_{\alpha \beta} \{ M_\beta \}$.

In the presence of a homogeneous external magnetic field $B_z {\bf e}_z$, the magnetization of the BEC precesses about the $z$ axis with the Larmor frequency. The time-averaged effective dipolar potential is obtained from the instantaneous form, Eq.~(\ref{appendixI}), by setting $K'_{xy}=0$ and replacing $K'_{xx}$ and $K'_{yy}$ with $\left(K'_{xx}+K'_{yy}\right)/2=-K'_{zz}/2$. Thus, in the special case of a condensate polarized in the $x$ direction, the time-averaged effective dipolar potential takes the simple form $\hat{V}_{\rm d}=-g_\mathrm{d} \hat{\mathcal{L}}_{zz} \{ M_x \} \hat{S}_x/2$.

With the notation introduced above, the two-dimensional time-dependent GPE can be restated as
\begin{eqnarray}
\label{appendix2DGP}
i\hbar \partial_t \Psi =\Big( \hat{h} &+& g_\mathrm{n} \Psi^\dagger \Psi + g_\mathrm{s} \Psi^\dagger \hat{S}_\alpha \Psi \hat{S}_\alpha \nonumber \\
&+& g_\mathrm{d} \hat{\mathcal{L}}_{\alpha \beta} \big\{ \Psi^\dagger \hat{S}_\beta \Psi \big\} \hat{S}_\alpha \Big) \Psi.
\end{eqnarray}
Direct substitution of the trial $\Psi(\br,t) = e^{-i\mu t/\hbar}\left[\Psi(\br) + \bu e^{-i\omega t}+\bv^\ast e^{i\omega^* t} \right]$ reveals that, in the absence of dipolar interactions, the linear operators $\hat{\mathcal{A}}$ and $\hat{\mathcal{B}}$ appearing in Eq.~(\ref{BdG}) can be written as
\begin{eqnarray}
\hat{\mathcal{A}}_{\rm ns} &=& \hat{h} +g_\mathrm{n} \left(\Psi^\dagger \Psi + \Psi \Psi^\dagger \right) \nonumber \\
& & + g_\mathrm{s} \left( \Psi^\dagger \hat{S}_\alpha \Psi \hat{S}_\alpha+\hat{S}_\alpha \Psi \Psi^\dagger \hat{S}_\alpha \right),\\
\hat{\mathcal{B}}_{\rm ns} &=& g_\mathrm{n} \Psi \Psi^T + g_\mathrm{s} \hat{S}_\alpha \Psi \Psi^T \hat{S}^*_\alpha.
\end{eqnarray}
When dipolar interactions are present, the operators $\hat{\mathcal{A}}$ and $\hat{\mathcal{B}}$ are augmented with the nonlocal terms 
\begin{eqnarray}
\hat{\mathcal{A}}_\mathrm{d} &=& g_\mathrm{d} \left( \hat{\mathcal{L}}_{\alpha \beta} \big\{ \Psi^\dagger \hat{S}_\beta \Psi \big\} \hat{S}_\alpha +
\hat{\mathcal{L}}_{\alpha \beta} \big\{ \Psi^\dagger \hat{S}_\beta \cdot \big\} \hat{S}_\alpha \Psi \right),\\
\hat{\mathcal{B}}_\mathrm{d} &=& g_\mathrm{d} \hat{\mathcal{L}}_{\alpha \beta} \big\{ \Psi^T \hat{S}^*_\beta \cdot \big\} \hat{S}_\alpha \Psi.
\end{eqnarray}
Therefore, in the general case, the operators appearing in Eq.~(\ref{BdG}) are given by $\hat{\mathcal{A}}=\hat{\mathcal{A}}_{\rm ns}+\hat{\mathcal{A}}_{\rm d}$ and $\hat{\mathcal{B}}=\hat{\mathcal{B}}_{\rm ns}+\hat{\mathcal{B}}_{\rm d}$.

\bibliography{JabRef}

\end{document}